\documentclass[12pt]{article}
\usepackage{latexsym}
\usepackage{seceqn}
\usepackage{epsf}
\usepackage{othersym}

\newcommand{\andvol}[3]{{\bf #1}~(#3)~#2}
\newcommand{\SCIENCE}[3]{Science~\andvol{#1}{#2}{#3}}
\newcommand{\PRL}[3]{Phys.~Rev.~Lett.~\andvol{#1}{#2}{#3}}
\newcommand{\PRD}[3]{Phys.~Rev.~\andvol{D#1}{#2}{#3}}

\newcommand{\NPB}[3]{Nucl.~Phys.~\andvol{B#1}{#2}{#3}}
\newcommand{\PLB}[3]{Phys.~Lett.~\andvol{B#1}{#2}{#3}}

\newcommand{\ZPC}[3]{Z.~Phys.~\andvol{C#1}{#2}{#3}}

\newcommand{\PTPS}[3]{Prog.~Theor.~Phys.~Suppl.~\andvol{#1}{#2}{#3}}
\newcommand{\MPLA}[3]{Mod.~Phys.~Lett.~\andvol{A#1}{#2}{#3}}

\newcommand{\EPJC}[3]{Euro.~Phys.~J.~\andvol{C#1}{#2}{#3}}
\newcommand{\hepph}[1]{\ hep-ph/#1}

\newcommand{\hepex}[1]{\ hep-ex/#1}

\newcommand{\etal}{\textit{et al.}}

\newcommand{\Zbb}{Zb\bar{b}}
\newcommand{\glb}{g_L^b}
\newcommand{\grb}{g_R^b}
\newcommand{\dglb}{\delta g_L^b}
\newcommand{\dgrb}{\delta g_R^b}
\newcommand{\ds}{\delta s^2}
\newcommand{\da}{\delta\alpha_s}
\newcommand{\drho}{\delta\rho}
\newcommand{\seff}{\sin^2\theta_\mathrm{eff}^\mathrm{lept}}
\newcommand{\sighad}{\sigma_\mathrm{had}^0}
\newcommand{\GamZ}{\Gamma_Z}
\newcommand{\Gamhad}{\Gamma_\mathrm{had}}
\newcommand{\Gamb}{\Gamma_{b\bar{b}}}
\newcommand{\Gamd}{\Gamma_{d\bar{d}}}

\newcommand{\Gaml}{\Gamma_{\ell^+\ell^-}}

\newcommand{\xib}{\xi_b}
\newcommand{\zetab}{\zeta_b}
\newcommand{\QWCS}{Q_W({}^{133}_{\phantom{1}55}{\rm Cs})}
\newcommand{\QWTL}{Q_W({}^{205}_{\phantom{1}81}{\rm Tl})}
\newcommand{\mt}{M_\mathrm{top}}
\newcommand{\mh}{M_\mathrm{higgs}}
\newcommand{\mz}{M_Z}
\newcommand{\mw}{M_W}
\newcommand{\SM}[1]{[ #1 ]_\mathrm{SM}}

\newcommand{\AQ}[1]{\Pi_{\rm #1}(q^2)}
\newcommand{\AZERO}[1]{\Pi_{\rm #1}(0)}

\newcommand{\APZERO}[1]{\Pi'_{\rm #1}(0)}

\newcommand{\beq}{\begin{equation}}
\newcommand{\eeq}{\end{equation}}
\newcommand{\beqa}{\begin{eqnarray}}
\newcommand{\eeqa}{\end{eqnarray}}
\newcommand{\beqx}{\begin{displaymath}}
\newcommand{\eeqx}{\end{displaymath}}
\newcommand{\beqax}{\begin{eqnarray*}}
\newcommand{\eeqax}{\end{eqnarray*}}
\begin{document}

\begin{titlepage}

\renewcommand{\thefootnote}{\fnsymbol{footnote}}

\begin{flushright}
VPI-IPPAP-98-1\\
UCLA/98/TEP/20\\
hep-ph/9807413\\
July 1998
\end{flushright}

\bigskip

\begin{center}

\textbf{\large
An analysis of Precision Electroweak Measurements:\\
Summer 1998 Update.
}

\bigskip
\bigskip

\textsc{Aaron~K.~GRANT}\footnote{%
email: grant@physics.ucla.edu,
Address from fall 1998: Department of Physics, Harvard University.}
\\
\textit{Department of Physics, University of California and Los Angeles,\\
 Los Angeles, CA 90095--1547}
\\
and
\\
\textsc{Tatsu~TAKEUCHI}\footnote{%
email: takeuchi@vt.edu}
\\
\textit{Institute for Particle Physics and Astrophysics\\
Physics Department, Virginia Tech, Blacksburg, VA 24061--0435}
\\

\bigskip
\bigskip

\begin{abstract}
We update our analysis of precision electroweak measurements
using the latest data announced at Moriond, March 1998.
Possible oblique corrections from new physics are parametrized using
the $STU$ formalism of Ref.~\cite{PESKIN:90}, and non--oblique
corrections to the $Zb\bar{b}$ vertex are parametrized
using the $\xi_b$--$\zeta_b$ formalism of Ref.~\cite{DPF:94}.
The implication of the analysis on minimal $SU(5)$ grand unification
is discussed.
\end{abstract}

\end{center}

\vfill

\begin{flushleft}
VPI-IPPAP-98-1\\
UCLA/98/TEP/20\\
hep-ph/9807413\\
July 1998
\end{flushleft}

\end{titlepage}

\renewcommand{\thefootnote}{\arabic{footnote}}
\setcounter{footnote}{0}

\section{Introduction}

The analysis of precision electroweak measurements provides us with
one of the few opportunities to constrain new physics beyond the
Standard Model.  The effectiveness of the approach is evident in the
prediction of the top quark mass which was predicted to be around 180
GeV \cite{Langacker:93} well before its direct measurement some time
later \cite{TOPMASS:95}.

In the past few years, we have seen a few notable developments
in the field of precision electroweak measurements.
In addition to the ever increasing accuracy of the LEP and SLD
measurements \cite{LEP:98}, a number of new or updated measurements
have been announced:
\begin{itemize}
\item{ 
The University of Colorado Group announced a new measurement of
the weak charge of Cesium~133 which improves the experimental error by
a factor of 7 compared to their previous measurement
\cite{WOOD:97,DZUBA:97}.   }

\item{ 
The CCFR/NuTeV collaboration has announced a preliminary
determination of $1-M_W^2/M_Z^2$ from $\nu N$ deep inelastic
scattering, \cite{NUTEV:98} which already improves on the previous
result from CCFR \cite{CCFR:98} by a factor of 2.  }

\item{ 
With the start of LEP2 and new analyses of data from CDF and
D0, the error on the $W$ mass has improved by more than a factor of 2
since the 1996 version of the Review of Particle
Properties. \cite{LEP2MW:98,PPBARMW:98} }

\item{ Of the measurements done by LEP at the $Z$ resonance, updated
values of $A^{0,b}_{FB}$ from ALEPH \cite{ALEPH:98} and $A_{\tau,e}$ from
L3 \cite{L3:98} are noteworthy, as they shift the
preferred value of $\seff$ somewhat.  }

\end{itemize}
In light of these developments, it is worthwhile to revisit these data
in hopes of assessing the status of the standard model and prospects
for new physics.

In this letter, we present the constraints imposed on new physics from
experimental data available as of June 1998.  In section 2, we restrict
our attention to oblique electroweak corrections and present the
results in terms of the $S$, $T$, $U$ parameters introduced in
Ref.~\cite{PESKIN:90}.  In section 3, we analyze the heavy flavor
observables from LEP and SLD for possible non--oblique corrections to
the $Zb\bar{b}$ vertex using the formalism of Ref.~\cite{DPF:94}.  In
this analysis, we let $\alpha_s(M_z)$ float and fit it to the data
also.
In section 4, we discuss the implications of our results for minimal
$SU(5)$ grand unification.  Section 5 concludes.

\section{Constraints on Oblique Electroweak Corrections}

The effects of new physics on electroweak observables can be quite
difficult to quantify.  Given the tremendous success of the standard
model in accounting for the data, however, it is reasonable to
restrict our attention by making some simplifying assumptions.  
This enables us to describe potential deviations from the
standard model in terms of just a few parameters.

The simplest, but not necessarily comprehensive, assumptions are
the following:
\begin{enumerate}
\item{The electroweak gauge group is the standard
      $SU(2)_L \times U(1)_Y$.  The only electroweak gauge
      bosons are the photon, the $W^\pm$, and the $Z$.}
\item{The couplings of new physics to light fermions are highly
      suppressed so that vertex and box corrections from new physics
      can be neglected (with the possible exception of processes
      involving the $b$ quark).  Only vacuum polarization
      (i.e. oblique) corrections need to be considered.  Further
      justifications of this approximation have been discussed in 
      Ref.~\cite{PESKIN:90} }
\item{The mass scale of new physics is large compared to the $W$
      and $Z$ masses.}
\end{enumerate}
These assumptions let us express the virtual effects of new physics
in terms of just three parameters defined as: \cite{PESKIN:90}
\beqa
\alpha S & = & 4s^2 c^2
               \left[ \APZERO{ZZ}
                      -\frac{c^2-s^2}{sc}\APZERO{Z\gamma}
                      -\APZERO{\gamma\gamma}
               \right]\,,  \nonumber \\
\alpha T & = & \frac{\AZERO{WW}}{\mw^2} - \frac{\AZERO{ZZ}}{\mz^2}\,, \\
\alpha U & = & 4s^2
               \left[ \APZERO{WW} - c^2\APZERO{ZZ}
                      - 2sc\APZERO{Z\gamma} - s^2\APZERO{\gamma\gamma} 
               \right]\,. \nonumber
               \phantom{\frac{e^2}{s^2}}   
\eeqa
Here, $\AQ{XY}$ is the transverse part of the vacuum polarization
function between gauge bosons $X$ and $Y$ and the prime represents a
derivative with respect to $q^2$.  $\alpha$ is the fine structure
constant and $s$ and $c$ are shorthand notations for the sine and
cosine of the weak mixing angle.  
Only the contribution of new physics to these functions are to be included.
The parameters $T$ and $U$ are defined so that
they vanish if new physics does not break the custodial $SU(2)$
symmetry. See Ref.~\cite{INAMI:92} for a discussion on the symmetry
properties of $S$.

The theoretical prediction for any observable will then consist of all
the standard model corrections to its tree level prediction plus the
possible corrections from new physics expressed in terms of $S$, $T$,
and $U$.  For instance, if the values of $\alpha$, $G_\mu$, and $\mz$
are used as input for the SM prediction, the shift in the $\rho$
parameter, the effective value of $\sin^2\theta_w$ in leptonic
asymmetries, and the $W$ mass due to new physics will be given by
\beqa
\rho -  \SM{\rho}
& = & \alpha T  \cr
\seff - \SM{\seff}
& = &   \frac{ \alpha }{ c^2 - s^2 }
        \left[ \frac{1}{4} S - s^2 c^2 T \right]  \cr
\mw/\SM{\mw}
& = & 1 + \frac{\alpha}{2(c^2-s^2)}
          \left[ -\frac{1}{2} S + c^2 T + \frac{c^2-s^2}{4 s^2} U
          \right],
\eeqa
where $\SM{\mathcal{O}}$ denotes the Standard Model prediction of
the observable $\mathcal{O}$.  

All other electroweak observables we will be considering get their
dependence on new physics corrections through $\rho$ and $\seff$.  As
a result, they only depend on $S$ and $T$, while the $W$ mass will be
the sole observable which depends on $U$.  By comparing standard model
predictions with experimental measurements, we can determine the
favored values of $S$, $T$, and $U$.  
The values of $S$, $T$ and $U$ obtained in this way give a
quantitative measure of the potential size of radiative corrections 
from new physics.  
If the standard model predictions for particular values of
$\alpha^{-1}(\mz)$, $\mt$ and $\mh$ yielded $S=T=U=0$, then
this would mean perfect agreement between the Standard Model and
experiment.  On the other hand, non-zero values of $S$, $T$ and $U$
would imply either that the experiments prefer the existence of
extra corrections from physics beyond the Standard Model, 
or that the values of $\alpha^{-1}(\mz)$, $\mt$ and $\mh$ chosen in
defining the ``reference'' Standard model were not optimal.

In table~1, we show the data we will be using to constrain
$S$, $T$, and $U$.  To the best of our knowledge, this is a
comprehensive set of all precision electroweak measurements that are
likely to have an impact on the analysis.  We have excluded all the
heavy flavor observables from the present analysis, since the impact
of new physics on these quantities cannot be fully parametrized using
$S$, $T$ and $U$.  We will return to the heavy flavor measurements in
the next section.  Some comments are in order:
\begin{itemize}
\item{The $\nu_\mu e$ -- $\bar{\nu}_\mu e$ scattering
parameters $g_V^{\nu e}$ and $g_A^{\nu e}$ are defined as
\beqax
g_V^{\nu e} & \equiv & 2 g_V^\nu g_V^e, \cr
g_A^{\nu e} & \equiv & 2 g_A^\nu g_A^e,
\eeqax
where $g_V^f$ and $g_A^f$ are the effective vector and
axial--vector couplings of the fermion $f$ to the $Z$.
At tree level, we have
\beqax
g_V^{\nu e} & = & \rho \left( -\frac{1}{2} + 2s^2 \right), \cr
g_A^{\nu e} & = & \rho \left( -\frac{1}{2} \right).
\eeqax
}
\item{
The $\nu_\mu N$ -- $\bar{\nu}_\mu N$ deep inelastic scattering parameters
$g_L^2$ and $g_R^2$ are defined as
\beqax
g_L^2 & \equiv & u_L^2 + d_L^2, \cr
g_R^2 & \equiv & u_R^2 + d_R^2,
\eeqax
where $q_L$ and $q_R$ are the effective left--handed and
right--handed couplings of the quark $q$ to the $Z$.
At tree level, they are equal to
\beqax
g_L^2 & = & \rho \left( \frac{1}{2} - s^2 + \frac{5}{9}s^4 \right), \cr
g_R^2 & = & \rho \left( \frac{5}{9} s^4 \right).
\eeqax
The quantity measured by NuTeV is a certain linear combination of
$u_{L/R}^2$ and $d_{L/R}^2$ which is roughly equal to
the Paschos--Wolfenstein parameter: \cite{PASCHOS:73}
\beqx
R^- = g_L^2 - g_R^2 = \rho \left( \frac{1}{2} - s^2 \right).
\eeqx
See Ref.~\cite{NUTEV:98} for details.
}
\item{
The weak charge of atomic nuclei measured in atomic parity violation
experiments is defined as \cite{BOUCHIAT:74}
\beqx
Q_W(Z,N) \equiv -2[ C_{1u} (2Z+N) + C_{1d} (Z+2N) ],
\eeqx
where $C_{1q}$ ($q=u,d$) are parameters in the
parity violating low energy effective Lagrangian:
\beqx
\mathcal{L}_\mathrm{PV}
= \frac{G_\mu}{\sqrt{2}}
  \sum_{q=u,d}
  \left[ C_{1q} (\bar{e}\gamma_\mu \gamma_5 e)
                (\bar{q}\gamma^\mu          q)
       + C_{2q} (\bar{e}\gamma_\mu          e)
                (\bar{q}\gamma^\mu \gamma_5 q)
  \right].
\eeqx
At tree level, we have
\beqx
Q_W(Z,N) = \rho [ Z (1-4s^2) - N ].
\eeqx
}
\item{ The value of $\seff$ from LEP is that derived from purely
leptonic asymmetries only. We include both the LEP and SLD
measurements in the fit with the quoted errors.  Another approach has
been taken in Ref.~\cite{ROSNER:98} }

\item{ 
The value of $\Gaml$ is that derived from the LEP $Z$ lineshape
variables $\GamZ$, $\sighad$, and $R_\ell = \Gamhad/\Gaml$.  Using
this one value in our analysis is equivalent to using all three with
correlations taken into account.  }

\item{ 
The value of the $W$ mass is the average of direct
determinations from LEP2 \cite{LEP2MW:98} and $p\bar{p}$ colliders
\cite{PPBARMW:98}.  }
\end{itemize}

\begin{table}[t]
\begin{center}
\begin{tabular}{|l|c|c|c|}
\hline
Observable    & SM prediction & Measured Value & Reference\\
\hline\hline
\underline{$\nu_\mu e$ and $\bar{\nu}_\mu e$ scattering} & & & \\ 
$g_V^{\nu e}$ & $-0.0365$     & $-0.041  \pm 0.015$  & \cite{PDG:98} \\
$g_A^{\nu e}$ & $-0.5065$     & $-0.507  \pm 0.014$  & \cite{PDG:98} \\
\hline
\underline{Atomic Parity Violation} & & & \\
$\QWCS$       & $-73.19$      & $-72.41  \pm 0.84$   & \cite{PDG:98} \\
$\QWTL$       & $-116.8$      & $-114.8  \pm 3.6$    & \cite{PDG:98} \\
\hline
\underline{$\nu_\mu N$ and $\bar{\nu}_\mu N$ DIS} & & & \\
$g_L^2$       & $0.3031$      & $0.3009  \pm 0.0028$ & \cite{PDG:98} \\
$g_R^2$       & $0.0304$      & $0.0328  \pm 0.0030$ & \cite{PDG:98} \\
NuTeV         & $0.2289$      & $0.2277  \pm 0.0022$ & \cite{NUTEV:98} \\
\hline
\underline{LEP/SLD} & & & \\
$\Gaml$       & $0.08392$ GeV & $0.08391 \pm 0.00010$ GeV & \cite{LEP:98} \\
$\seff$ (LEP) & $0.23200$     & $0.23157 \pm 0.00041$     & \cite{LEP:98} \\
$\seff$ (SLD) & $0.23200$     & $0.23084 \pm 0.00035$     & \cite{LEP:98} \\
\hline
\underline{$W$ mass} & & & \\
$\mw$         & $80.315$ GeV  & $80.375 \pm 0.064$ GeV    & 
\cite{LEP2MW:98,PPBARMW:98} \\
\hline
\end{tabular}
\caption{%
The data used for the oblique correction analysis.  The value
of $\seff$ for LEP is from leptonic asymmetries only.  The $W$ mass is
the average of LEP2 \cite{LEP2MW:98} and $p\bar{p}$ \cite{PPBARMW:98}
values.  Definitions of $g_{V,A}^{\nu e}$ and $g_{L,R}^2$ can be found
in the Review of Particle Properties \cite{PDG:98}.  The SM
predictions for the $W$ mass and the LEP/SLC observables were obtained
using the program ZFITTER 4.9 \cite{ZFITTER:92}, and the predictions
for the other low energy observables were calculated from the formulae
given in Ref.~\cite{MARCIANO:80}.  The parameter choice for the
reference SM was $\mz = 91.1867$ GeV \cite{LEP:98}, $\mt = 173.9$ GeV
\cite{TOPMASS:98}, $\mh = 300$ GeV, $\alpha^{-1}(\mz) = 128.9$
\cite{ALEMANY:98}, and $\alpha_s(\mz) = 0.120$. }
\end{center}
\end{table}

To fix the reference Standard Model to which we compare the
experimental data, we use the values
\cite{TOPMASS:98,MHIGGS:98,ALEMANY:98}
\beqa
\mt & = & 173.9\,\mathrm{GeV},\cr
\mh & = & 300\,\mathrm{GeV},\cr
\alpha^{-1}(\mz) & = & 128.9, \cr
\alpha_s(\mz) & = & 0.12
\eeqa
Fitting to the data of table~1, we find
\beqa
S & = & -0.33 \pm 0.14 \cr
T & = & -0.14 \pm 0.15 \cr
U & = & \phantom{-}0.07 \pm 0.22 
\eeqa
with the correlation matrix given by
\beq
\left[ \begin{array}{lll}
       \phantom{-}1     &  \phantom{-}0.85  & -0.21 \\
       \phantom{-}0.85  &  \phantom{-}1     & -0.42 \\
       -0.21            &  -0.42            & \phantom{-}1 
       \end{array}
\right]
\eeq
where rows and columns are labelled in the order $S$, $T$, $U$.  The
quality of the fit is $\chi^2 = 4.5/(11-3)$.  Compared to the results
of the 1996 \cite{HEWETT:96} data, the major improvement is in the
limits on $U$: the error has been reduced by more than a factor of 2.  
This can be directly traced to the improvement in the value of
$\mw$.

\begin{figure}[p]
\centering
\unitlength=1cm

\begin{picture}(15,10)
\unitlength=1mm
\put(55,48){$g_V^{\nu e}$}
\put(117,45){$g_A^{\nu e}$}
\epsfbox[69 69 470 350]{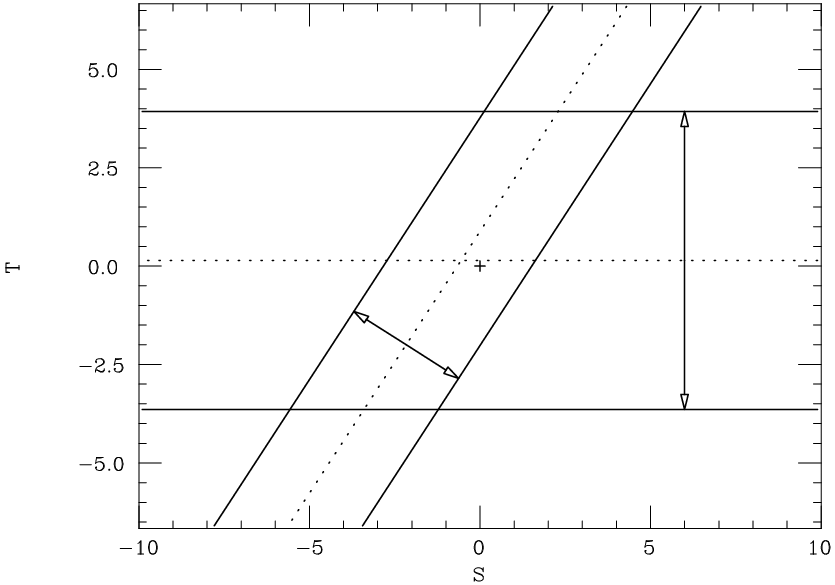}
\end{picture}

\vspace{5pt}

\begin{picture}(15,10)
\unitlength=1mm
\put(96,70){68\%}
\put(110,56){90\%}
\epsfbox[69 69 470 350]{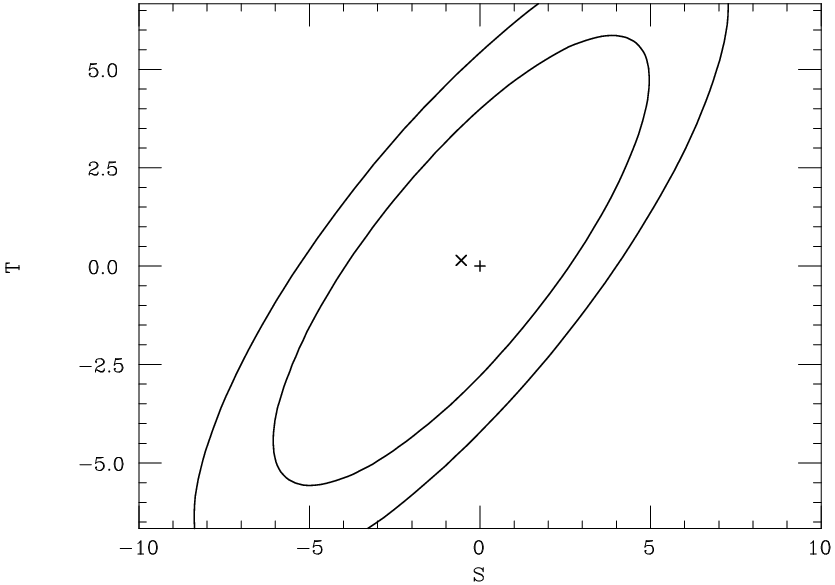}
\end{picture}

\vspace{-10pt}
\caption[]{%
The limits on $S$ and $T$ from $\nu_\mu e$ and
$\bar{\nu}_\mu e$ scattering experiments.  }
\label{FIGENU}
\end{figure}

\begin{figure}[p]
\centering
\unitlength=1cm

\begin{picture}(15,10)
\unitlength=1mm
\put(54,33){$Q_w$(Cs)}
\put(35,72){$Q_w$(Tl)}
\epsfbox[69 69 470 350]{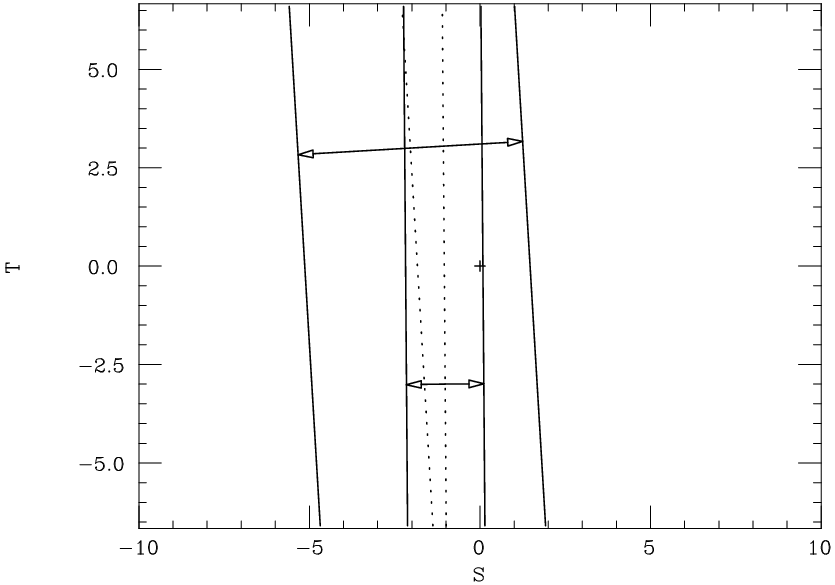}
\end{picture}

\vspace{5pt}

\begin{picture}(15,10)
\unitlength=1mm
\put(74,80){68\%}
\put(89,80){90\%}
\epsfbox[69 69 470 350]{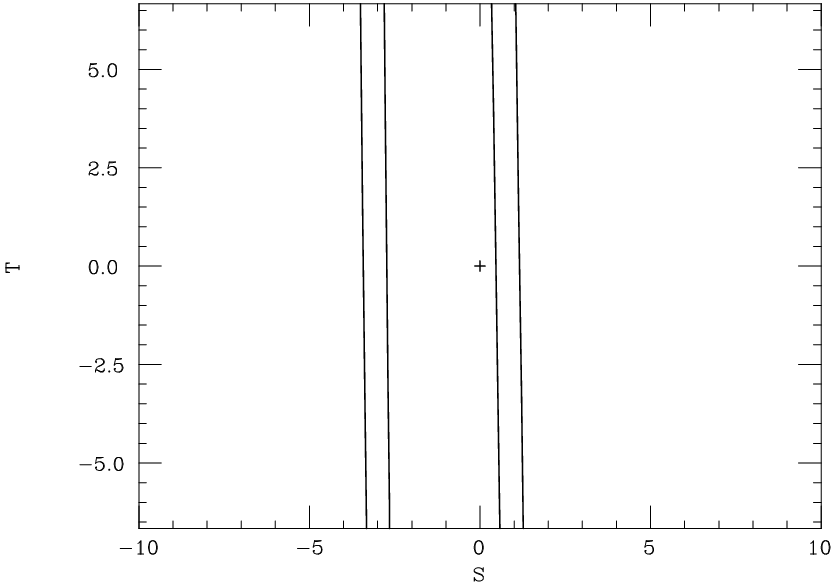}
\end{picture}

\vspace{-10pt}
\caption[]{%
The limits on $S$ and $T$ from atomic parity violation
experiments.  }
\label{FIGAPV}
\end{figure}

\begin{figure}[p]
\centering
\unitlength=1cm

\begin{picture}(15,10)
\unitlength=1mm
\put(126,66){$g_L^2$}
\put(111,25){$g_R^2$}
\put(46,22){NuTeV}
\epsfbox[69 69 470 350]{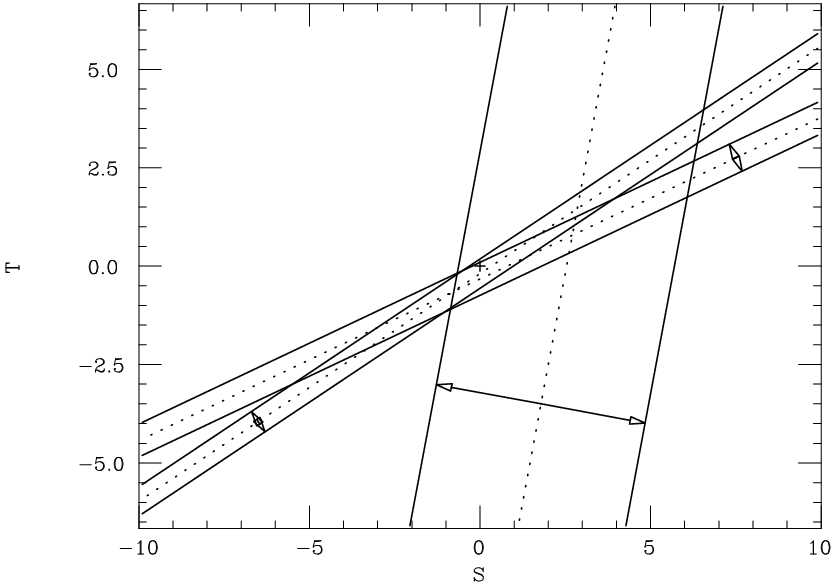}
\end{picture}

\vspace{5pt}

\begin{picture}(15,10)
\unitlength=1mm
\put(44,63){68\%}
\put(44,52){90\%}
\put(118,60){without}
\put(118,55){NuTeV}
\epsfbox[69 69 470 350]{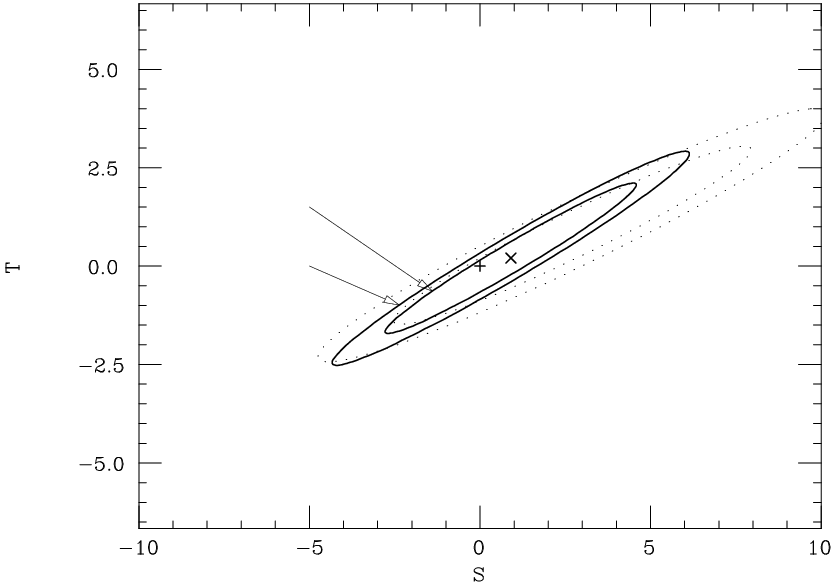}
\end{picture}

\vspace{-10pt}
\caption[]{%
The limits on $S$ and $T$ from $\nu_\mu N$ and
$\bar{\nu}_\mu N$ deep inelastic scattering experiments.  }
\label{FIGDIS}
\end{figure}

\begin{figure}[p]
\centering
\unitlength=1cm

\begin{picture}(15,10)
\unitlength=1mm
\put(120,79){$M_W$}
\put(123,51){$\Gamma_{\ell^+\ell^-}$}
\put(66,84){SLD}
\put(62,18){LEP leptonic asymmetries}
\epsfbox[69 69 470 350]{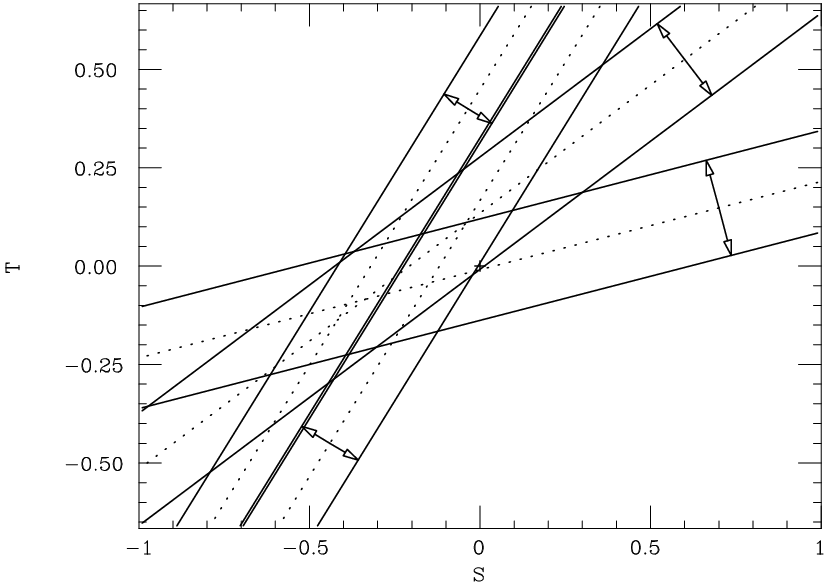}
\end{picture}

\vspace{5pt}

\begin{picture}(15,10)
\unitlength=1mm
\put(55,40){68\%}
\put(42,40){90\%}
\epsfbox[69 69 470 350]{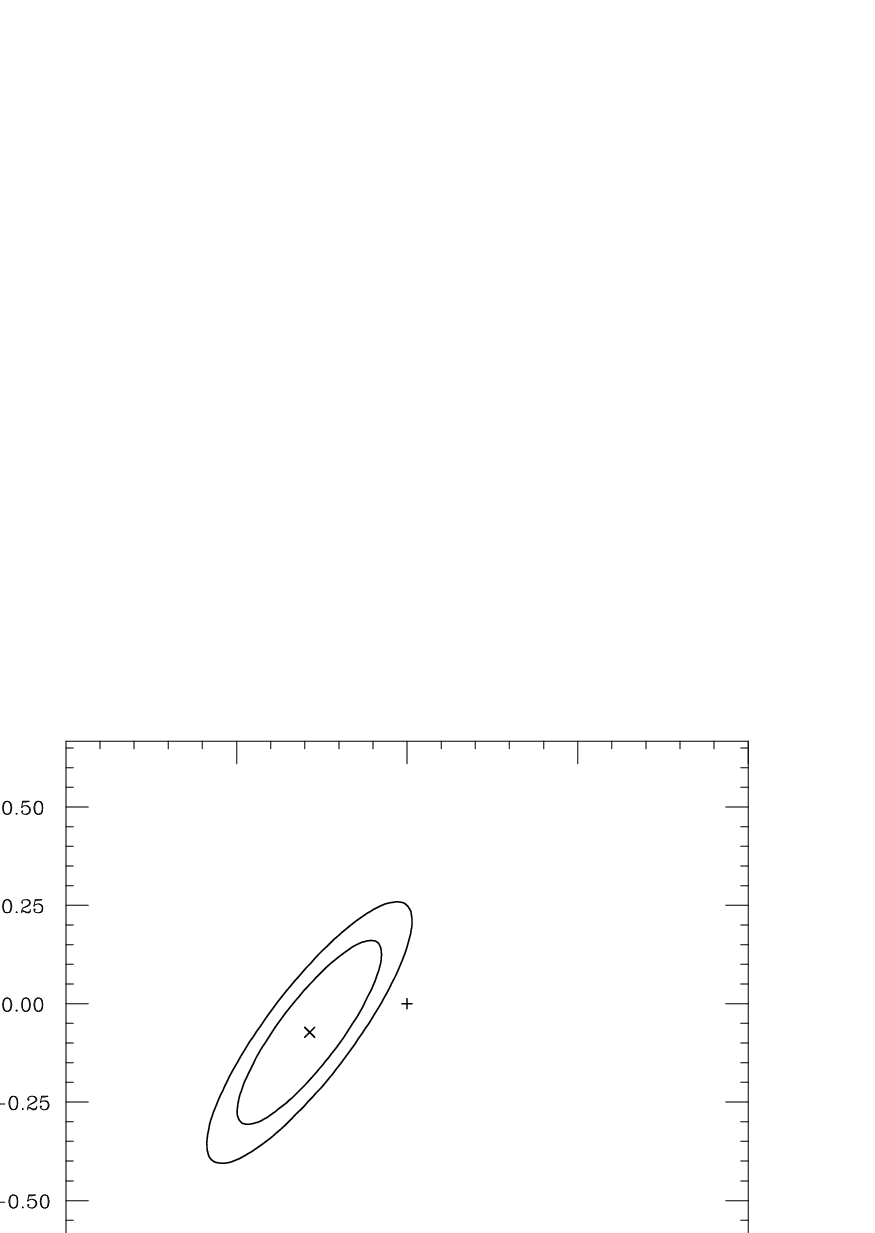}
\end{picture}

\vspace{-10pt}
\caption[]{%
The limits on $S$ and $T$ from SLD, LEP, and
$M_W$ (LEP2 and $p\bar{p}$).
}
\label{FIGLEP}
\end{figure}

\begin{figure}[tp]
\centering
\unitlength=1cm

\begin{picture}(15,10)
\unitlength=1mm
\epsfbox[69 69 470 350]{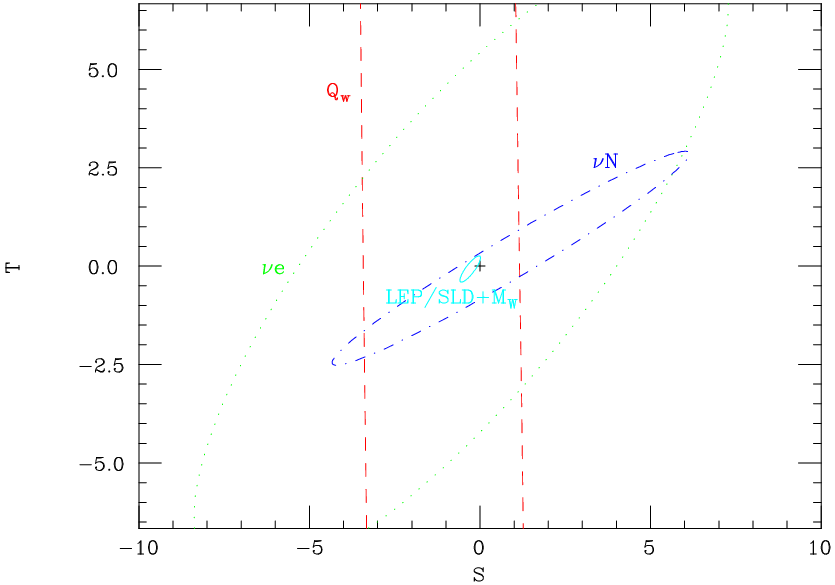}
\end{picture}

\vspace{-10pt}
\caption[]{%
Comparison of
the 90\% likelihood contours from different experiments.
}
\label{FIG90}


\begin{picture}(15,10)
\unitlength=1mm
\put(54,38){68\%}
\put(40,38){90\%}
\epsfbox[69 69 470 350]{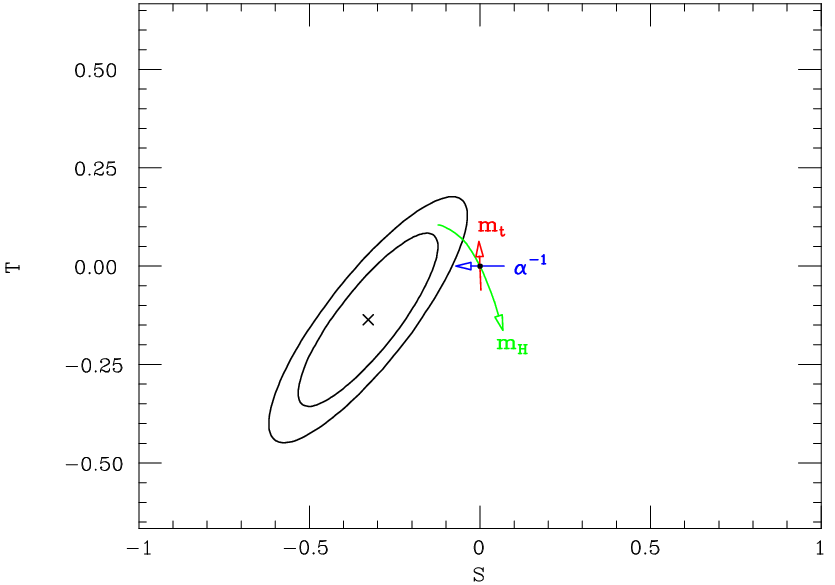}
\end{picture}

\vspace{-10pt}
\caption[]{%
The limits on $S$ and $T$ : all experiments combined.
The arrows show the range the SM point will move when
$\mt$, $\mh$, and $\alpha^{-1}(\mz)$ are varied.
Red arrow:
$\mt$ varied from 168.7 to 179.1~GeV \cite{TOPMASS:98},
Green arrow:
$\mh$ varied from 80 to 1000~GeV \cite{MHIGGS:98},
Blue arrow:
$\alpha^{-1}(\mz)$ varied from 128.8 to 129.0 \cite{ALEMANY:98}.
}
\label{FIGALL}
\end{figure}

In Figs.~\ref{FIGENU} to \ref{FIGLEP}, we show the limits placed on
$S$ and $T$ separately by each class of experiments.  
The bands in the upper figures represent the 1--$\sigma$ limits
placed on $S$ and $T$ by each observable.
Note that there is an overall change in scale between 
Figs.~\ref{FIGDIS} and \ref{FIGLEP}.  
In Fig.~\ref{FIG90}, we compare the 90\% confidence
limits placed on $S$ and $T$ by the four classes of experiments we
consider, while Fig.~\ref{FIGALL} shows the limits on $S$ and $T$
combining all experiments.

We can see from Fig.~\ref{FIGALL} that the current data favor
either a small value of the Higgs mass or a larger value of
$\alpha^{-1}(\mz)$.  The indications of a light Higgs would be consistent
with low energy supersymmetry, which predicts a Higgs lighter than
about 130 GeV \cite{CHANKOWSKI:97}, and typically gives small
contributions to the oblique parameters \cite{PIERCE:98}.


%
%
%
%


\newpage

\section{Constraints on Non-Oblique Corrections to the $\Zbb$ Vertex}

In the previous section we excluded heavy flavor observables from our
analysis, since in principle there could be corrections to these
quantities that cannot be described solely in terms of $S,~T$ and $U$.
In this section, we extend our analysis to include heavy flavor
observables.  This of course entails additional assumptions beyond
those enumerated at the beginning of Sec.~2: in particular, we now
assume:

\begin{enumerate}
\item{	The couplings of light ($u,d,s,c$) quarks to the $Z$ are
	dictated solely by the standard model together with possible oblique
	corrections from new physics.}
\item{  The couplings of the $b$ to the $Z$ may exhibit additional
	deviations from the standard model in the form of ``direct''
	or ``non-oblique'' corrections; that is to say, the couplings
	of the $b$ may receive appreciable corrections from vertex
	diagrams in addition to corrections from vacuum polarization
	diagrams.}
\end{enumerate}

These additional assumptions may appear on the face of it to be quite
artificial, and indeed they do restrict considerably the class of
models that are accurately described by our analysis.  Just the same,
however, these assumptions are valid for a large class of models.  The
reason is that the $b$ is the isospin partner of the top, and hence
its couplings can be modified by the mechanism responsible for generating
the large top mass.
Indeed, even in the standard model, the $b$ receives ``non-oblique''
corrections that are absent for the first two generations.
Appreciable non-oblique corrections to the $b$ couplings would be
expected generically in models with extended Higgs sectors and in
models where the $(t,b)$ doublet is involved directly in electroweak
symmetry breaking.

\begin{figure}[tp]
\centering
\unitlength=1cm

\begin{picture}(15,10)
\unitlength=1mm
\epsfbox[12 80 408 354]{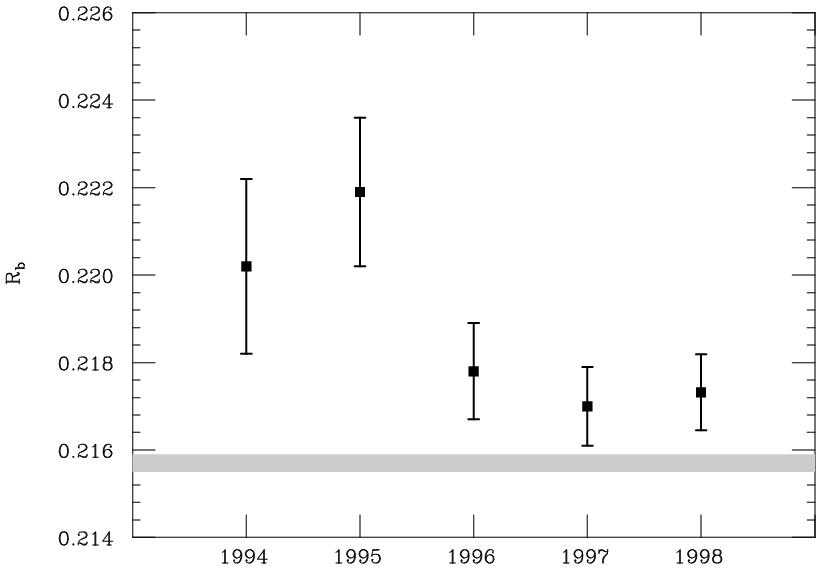}
\end{picture}
\vspace{-10pt}
\caption[]{
Change in the experimental value of $R_b$.
The Standard Model prediction is shown by the shaded band.}
\label{figRb}

%

\begin{picture}(15,10)
\unitlength=1mm
\epsfbox[12 80 408 354]{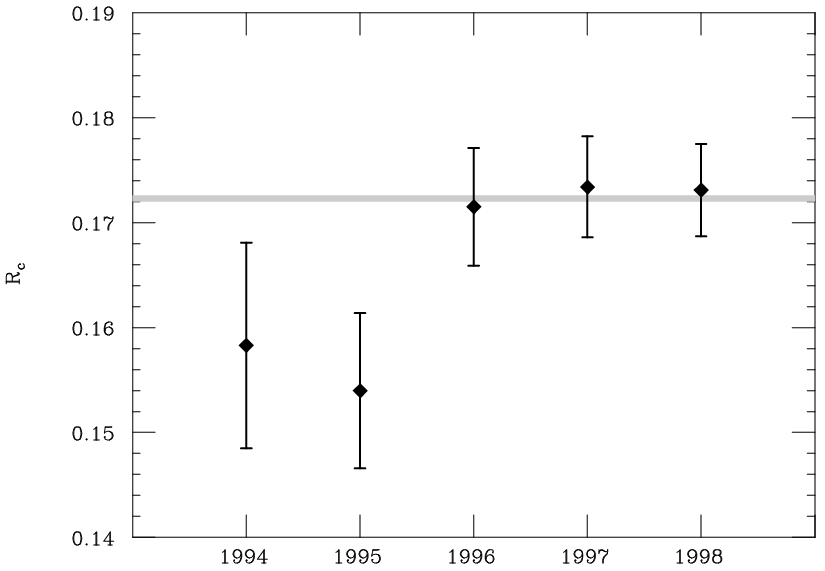}
\end{picture}
\vspace{-10pt}
\caption[]{%
Change in the experimental value of $R_c$.
The Standard Model prediction is shown by the shaded band.}
\label{figRc}
\end{figure}

Measurements of heavy flavor observables have shifted somewhat in
recent years as experimental understanding has
improved. \cite{LEP:OLD} In particular, an apparent excess in the
partial width of the $Z$ to $b$ quarks has decreased substantially,
improving the comparison between the standard model and
experiment. (cf. Figs.~\ref{figRb} and \ref{figRc}.)  In this section
we use the latest data \cite{LEP:98} to determine how well the
standard model describes the couplings of the $b$ to the $Z$.

We begin by defining
\beqa
\drho & = & \alpha T,    \cr
\ds   & = & \frac{\alpha}{c^2-s^2}\left[ \frac{1}{4} S - s^2 c^2 T
                                  \right].
\eeqa
$\drho$ and $\ds$ are just the shifts of the $\rho$ parameter and
$\seff$:
\beqa
\rho  & = & [\rho]_{\rm SM} + \drho,   \cr
\seff & = & [\seff]_{\rm SM} + \ds.
\eeqa
We write the
left and right handed couplings of the $b$ quark to the $Z$ as
\beqa
\glb  =  [\glb]_{\rm SM} + \frac{1}{3}\,\ds + \dglb, \cr
\grb  =  [\grb]_{\rm SM} + \frac{1}{3}\,\ds + \dgrb,            
\eeqa
where we have included possible non--oblique corrections from 
new physics, $\dglb$ and $\dgrb$.  
Assuming that only the couplings of the $b$ are significantly affected by 
non--oblique corrections, we can compute the
dependence of electroweak observables on $\drho$, $\ds$, $\dglb$, and $\dgrb$.

It is convenient to define the following linear combinations
of $\dglb$ and $\dgrb$:
\beqa
\xib   & \equiv & (\cos\phi_b)\dglb - (\sin\phi_b)\dgrb,  \cr
\zetab & \equiv & (\sin\phi_b)\dglb + (\cos\phi_b)\dgrb,
\label{xizetadef}
\eeqa
where 
\beq
\phi_b \equiv \tan^{-1}|\grb/\glb| \approx 0.181.
\eeq
By expanding $\Gamb$ 
about the point $\ds=\xib=\zetab=0$,
we find
\beqa
\Gamb 
& = & [ \Gamb ]_{\rm SM} 
\left\{ 1 + \drho
          + \frac{2}{3}\left[ \frac{ \glb   + \grb }
                                   { (\glb)^2 + (\grb)^2 } 
                       \right] \ds
\right.
\nonumber \\
&   &  \quad\quad
\left.
          + \left[ \frac{ 2 }{ (\glb)^2 + (\grb)^2 }
            \right] 
            \left( \glb\,\dglb + \grb\,\dgrb
            \right)
\right\}
\nonumber \\
& = & 
[ \Gamb ]_{\rm SM} 
\left(  1 + \drho + 1.25\,\ds  - 4.65\,\xib
\right)
\eeqa
Similarly,
\beqa
A_b
& = & \frac{ (\glb)^2 - (\grb)^2 }{ (\glb)^2 + (\grb)^2 }    \cr
& = & [ A_b ]_{\rm SM}
      \left\{ 1 - \frac{ 4 }{ 3 }
                  \left[ \frac{ \glb\grb(\glb-\grb) }
                              { (\glb)^4 - (\grb)^4)   }
                  \right] \ds
      \right.  \cr
&   & \qquad\qquad
      \left.
                - \left[ \frac{ 4\glb\grb     }
                              { (\glb)^4 - (\grb)^4 }
                  \right]
                  \left( \glb\,\dgrb - \grb\,\dglb
                  \right)
      \right\}  \cr
& = & [ A_b ]_{\rm SM}
      \left( 1 - 0.68\,\ds - 1.76\,\zetab
      \right).
\label{Abzeta}
\eeqa
All the other observables get their dependence on
$\dglb$ and $\dgrb$ through either $\Gamb$ or
$A_b$ so they will depend on either $\xib$ or $\zetab$, but not both.
The observables that depend on $\Gamb$ are:
\beqa
\Gamma_Z & = & [\Gamma_Z]_{\rm SM}
               \left( 1 + \drho - 1.06\,\ds - 0.71\,\xib + 0.21\,\da 
               \right),   \cr
\sighad & = & [\sighad]_{\rm SM}
               \left( 1 + 0.11\,\ds + 0.41\,\xib - 0.12\,\da
               \right),   \cr
R_\ell \;\equiv\; \Gamma_{\rm had}/\Gamma_{\ell^+\ell^-}
    & = & [R_Z]_{\rm SM}
               \left( 1 - 0.85\,\ds - 1.02\,\xib + 0.31\,\da
               \right),   \cr
R_b \;\equiv\; \Gamma_{b\bar{b}}/\Gamma_{\rm had}
    & = & [R_b]_{\rm SM}
               \left( 1 + 0.18\,\ds - 3.63\,\xib
               \right),   \cr
R_c \;\equiv\; \Gamma_{c\bar{c}}/\Gamma_{\rm had}
    & = & [R_c]_{\rm SM}
               \left( 1 - 0.35\,\ds + 1.02\,\xib
               \right).
\label{Rxidep}
\eeqa
The parameter $\da$ is a possible shift of $\alpha_s(\mz)$ from
our reference value of 0.120,
\beqx
\alpha_s(\mz) = 0.120 + \da.
\eeqx
Note that only $\Gamma_Z$ depends on $\drho$.  We will ignore
$\Gamma_Z$ in the following since including it will only place limits
on $\drho$ without affecting the other parameters.  We will also
omit all non--LEP/SLD observables since these are expected to have a
negligible impact on the $b$ couplings.

In an analogous way, we find
\beq
A_{\rm FB}^{b,0} = \frac{3}{4} A_e A_b
                 = [A_{\rm FB}^b]_{\rm SM}
                   \left( 1 - 55.7\,\ds - 1.76\,\zetab
                   \right).
\label{AFBbzeta}
\eeq
The value of $A_{\rm FB}^{b,0}$ is the measured forward--backward
asymmetry of the $b$ with QCD corrections removed, so it naturally
depends on the value of $\alpha_s(\mz)$ used in the calculation.
Since we let the value of $\alpha_s(\mz)$ float in our fit, this
should be taken into account.  However, the dependence of the
extracted value of $A_{\rm FB}^{b,0}$ on $\alpha_s(\mz)$ is not
straightforward since it depends on the details of each LEP detector.
We estimated the sensitivity to $\alpha_s(\mz)$ using the formulae in
Ref.~\cite{ABBANEO:98} and found it to be negligibly small as long as
$|\da| \alt 0.1$.  The sensitivity to $\alpha_s(\mz)$ is smaller than
the systematic error ascribed to $A_{\rm FB}^{b,0}$. We will therefore
ignore the $\alpha_s(\mz)$ dependence of $A_{\rm FB}^{b,0}$
in our analysis, and similarly that of
$A_{\rm FB}^{c,0}$.

The relationship between our parameters and others that have appeared
in the literature is as follows.  The parameter $\epsilon_b$
introduced in Ref.~\cite{ALTARELLI:93} was defined as
\beqa
\glb - \grb & \equiv &  -\frac{1}{2}
              \left( 1 + \epsilon_b
              \right), \cr
\glb + \grb & \equiv &  -\frac{1}{2}
              \left( 1 - \frac{4}{3}\seff + \epsilon_b
              \right). 
\eeqa
This definition assumes $\dgrb = 0$, and the relation 
between $\epsilon_b$ and $\dglb$ is given by
\beq
\epsilon_b = [\epsilon_b]_{\rm SM} - 2\dglb.
\eeq
The parameters $\delta_{bV}$ and $\eta_b$ introduced in 
Ref.~\cite{BLONDEL:92} were defined as
\beqa
\Gamb & \equiv & \Gamd \,( 1 + \delta_{bV} ), \cr
A_b   & \equiv & A_s   \,( 1 + \eta_b ).
\eeqa
They are related to $\xib$ and $\zetab$ by
\beqa
\delta_{bV} & = & [\delta_{bV}]_{\rm SM} - 4.65\,\xib,  \cr
\eta_b      & = & [\eta_b]_{\rm SM}      - 1.76\,\zetab.
\eeqa

\begin{table}[t]
\begin{center}
\begin{tabular}{|l|c|c|}
\hline
Observable    & ZFITTER prediction & Measured Value \\
\hline\hline
$\seff$ (LEP) & $0.23200$     & $0.23157 \pm 0.00041$     \\
$\seff$ (SLD) & $0.23200$     & $0.23084 \pm 0.00035$     \\
$\sighad$     & $41.468$~nb   & $41.486  \pm 0.053$~nb    \\
$R_\ell$      & $20.749$      & $20.775  \pm 0.027$       \\
$R_b$         & $0.21575$     & $0.21732 \pm 0.00087$     \\
$R_c$         & $0.1723$      & $0.1731  \pm 0.0044$      \\
$A_\mathrm{FB}^{b,0}$
              & $0.1004$      & $0.0998  \pm 0.0022$      \\
$A_\mathrm{FB}^{c,0}$
              & $0.0716$      & $0.0735  \pm 0.0045$      \\
$A_b$         & $0.934$       & $0.899   \pm 0.049$       \\
$A_c$         & $0.666$       & $0.660   \pm 0.064$       \\
\hline
\end{tabular}
\caption{The data used for the $\Zbb$ vertex correction analysis.  All
data are from Ref.~\cite{LEP:98}.  The value of $\seff$ for LEP is
from leptonic asymmetries only.  The parameter choice for the
reference SM was $\mz = 91.1867$ GeV \cite{LEP:98}, $\mt = 173.9$ GeV
\cite{TOPMASS:98}, $\mh = 300$ GeV, $\alpha^{-1}(\mz) = 128.9$
\cite{ALEMANY:98}, and $\alpha_s(\mz) = 0.120$.}
\end{center}
\end{table}

In table~2, we show the data used in our analysis.  A fit
to this data with $\ds$, $\xib$, $\zetab$, and $\da$ as parameters,
including the correlations between $\sighad$ and $R_\ell$, and among
all heavy flavor observables yields:
\beqa
\ds    & = & -0.00082 \pm 0.00025,  \cr
\xib   & = & -0.0021  \pm 0.0011,   \cr
\zetab & = & \phantom{-}0.028 \pm 0.013, \cr
\da    & = & -0.006   \pm 0.005
\eeqa
with the correlation matrix given by
\beq
\left[ 
\begin{array}{llll}
\phantom{-}1    & \phantom{-}0.02 & -0.49           & \phantom{-}0.13 \\
\phantom{-}0.02 & \phantom{-}1    & -0.02           & \phantom{-}0.70 \\
-0.49           & -0.02           & \phantom{-}1    & -0.07           \\
\phantom{-}0.13 & \phantom{-}0.70 & -0.07           & \phantom{-}1    
\end{array}
\right]
\eeq
where the rows and columns are labelled in the order $(\ds, \xib,
\zetab, \da)$.  The quality of the fit was $\chi^2 = 2.5/(10-4)$.

\begin{figure}[tp]
\centering
\unitlength=1cm

\begin{picture}(15,10)
\unitlength=1mm
\put(36,43){$R_b$}
\put(101.5,80){$R_c$}
\put(62,81){$A_\mathrm{FB}^{0,c}$}
\put(61,28){SLD}
\put(80,28){LEP}
\put(80,24){leptonic}
\put(80,20){asymmetries}
\put(116,33){$R_\ell$}
\put(120,63){$\sigma_h^0$}
\epsfbox[16 64 412 356]{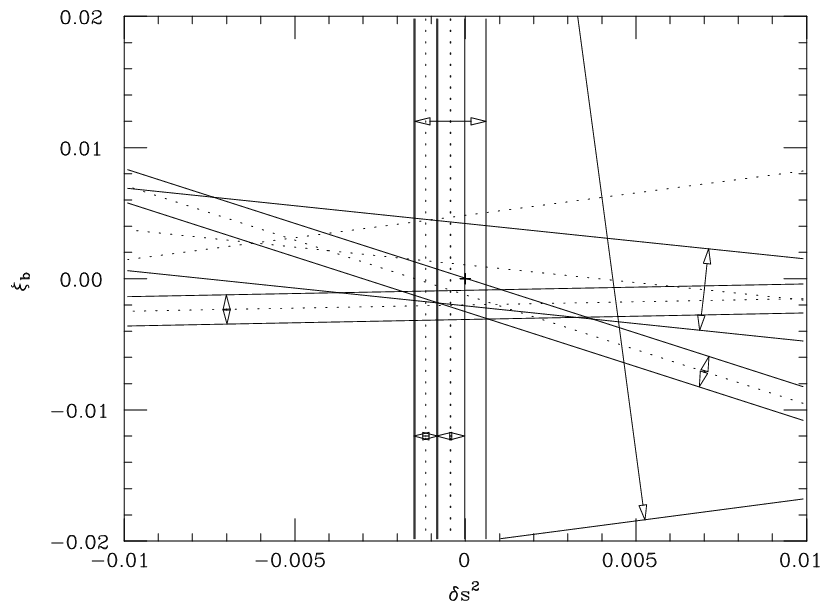}
\end{picture}

\vspace{-10pt}
\caption[]{%
1--$\sigma$ limits on $\delta s^2$ and $\xi_b$.
}
\label{fig1a}

%

\vspace{10pt}

\begin{picture}(15,10)
\unitlength=1mm
\put(36,50){$A_b$}
\put(104,30){$A_\mathrm{FB}^{0,b}$}
\put(83,81){$A_\mathrm{FB}^{0,c}$}
\put(61,28){SLD}
\put(80,28){LEP}
\put(80,24){leptonic}
\put(80,20){asymmetries}
\epsfbox[16 64 412 356]{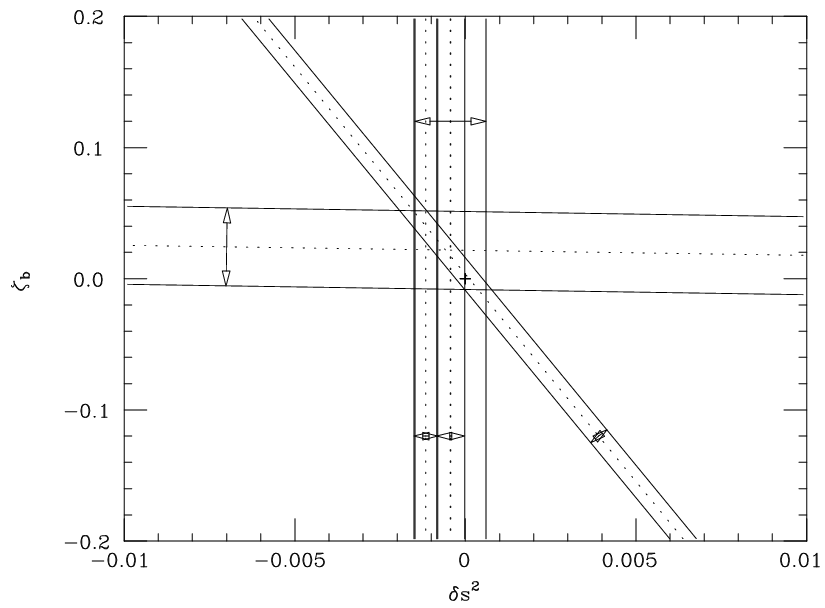}
\end{picture}

\vspace{-10pt}
\caption[]{%
1--$\sigma$ limits on $\delta s^2$ and $\zeta_b$.
}
\label{fig1b}
\end{figure}

\begin{figure}[tp]
\centering
\unitlength=1cm

\begin{picture}(15,10)
\unitlength=1mm
\put(94,57){Standard Model}
\put(54,28){68\%}
\put(39,28){90\%}
\epsfbox[16 64 412 356]{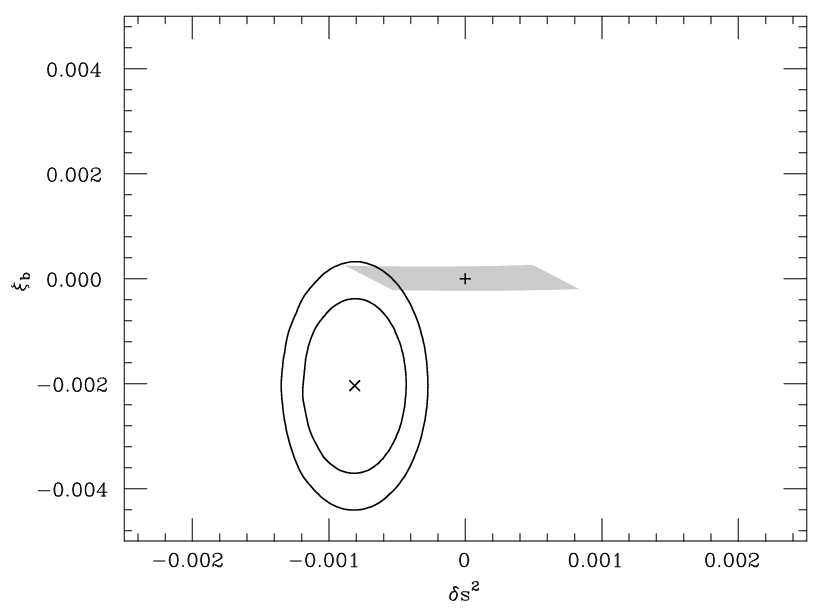}
\end{picture}

\vspace{-10pt}
\caption[]{%
Limits on $\delta s^2$ and $\xi_b$.
The shaded area represents the Standard Model points
with $\mt=168.7\sim 179.1$ GeV, and $\mh=80\sim 1000$GeV. 
}
\label{fig2a}

\vspace{10pt}

%

\begin{picture}(15,10)
\unitlength=1mm
\put(94,31){Standard Model}
\put(53,62){68\%}
\put(38,62){90\%}
\epsfbox[16 64 412 356]{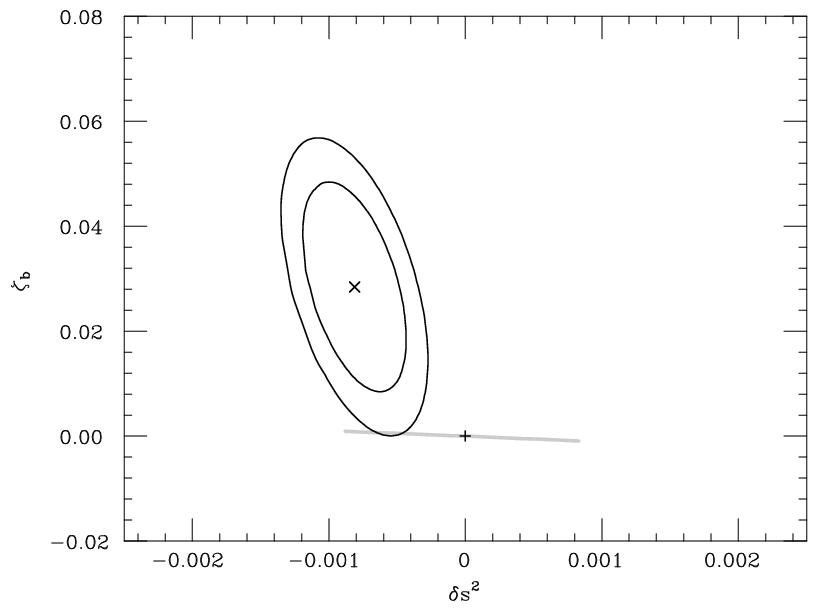}
\end{picture}

\vspace{-10pt}
\caption[]{%
Limits on $\delta s^2$ and $\zeta_b$.
The shaded area represents the Standard Model points
with $\mt=168.7\sim 179.1$ GeV, and $\mh=80\sim 1000$GeV. 
}
\label{fig2b}
\end{figure}

The constraints imposed on $\ds$, $\xib$, and $\zetab$ by the various
observables are illustrated in Figs~\ref{fig1a} through \ref{fig2b}.
The bands in Figs.~\ref{fig1a} and \ref{fig1b} illustrate the
$1-\sigma$ uncertainties on the various constraints.  The
2--dimensional projections of the allowed regions onto the
$\ds$--$\xib$, $\ds$--$\zetab$ planes are shown in Figs.~\ref{fig2a}
and \ref{fig2b}.

\begin{figure}[tp]
\centering
\unitlength=1cm

\begin{picture}(15,13)
\unitlength=1mm
\put(76,33){\Large $\xi_b$}
\put(56,64){\Large $\zeta_b$}
\put(87,48){Standard Model}
\epsfbox[50 70 452 436]{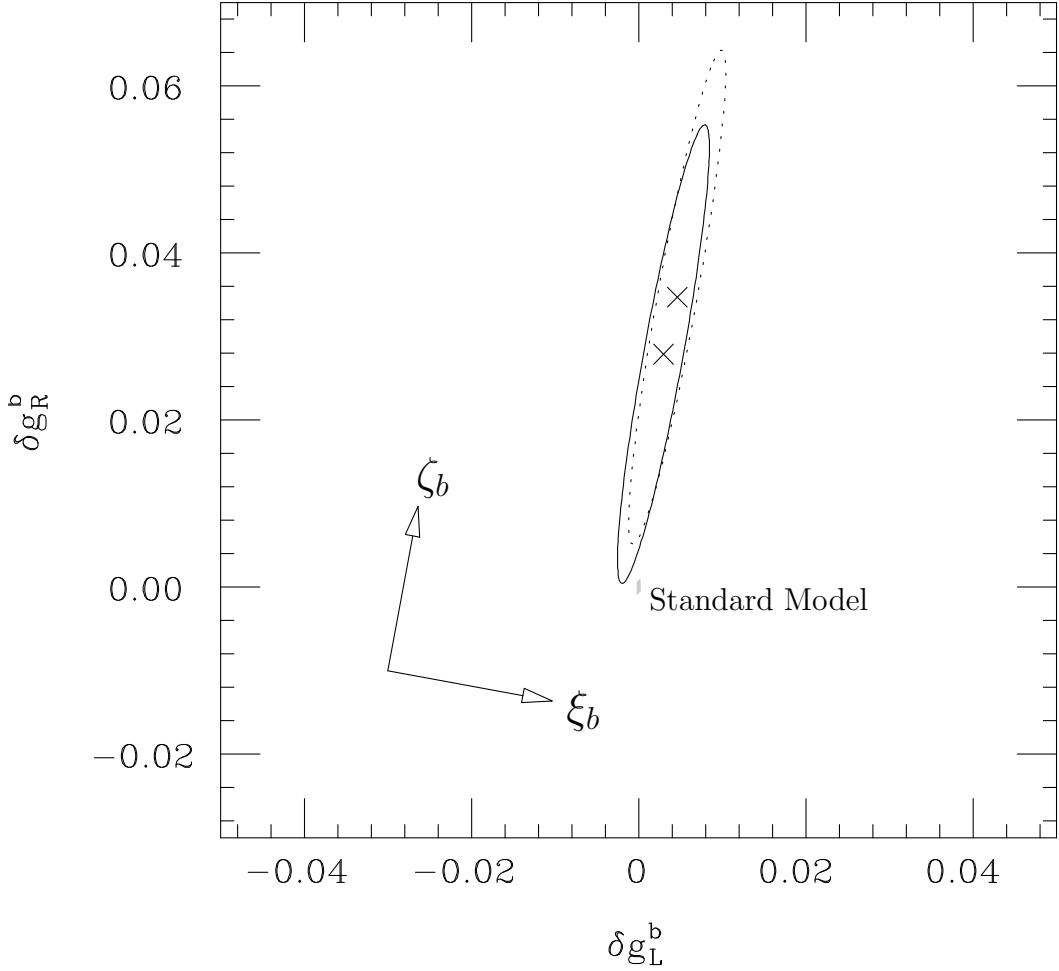}
\end{picture}

\vspace{-10pt}
\caption[]{%
90\% confidence limits on $g_L^b$ and $g_R^b$.
Solid line -- 98 data, dotted line -- 97 data.
The small shaded area around the origin represents the Standard Model points
with $\mt=168.7\sim 179.1$ GeV, and $\mh=80\sim 1000$GeV. 
}
\label{fig3}
\end{figure}

In terms of $\dglb$ and $\dgrb$, the limits on
$\xib$ and $\zetab$ translate into
\beqa
\label{couplings}
\dglb  & = &  0.0030 \pm 0.0026,           \cr
\dgrb  & = &  0.028\phantom{0} \pm 0.013.  \cr
{\rm corr}(\dglb, \dgrb) & = & 0.9
\eeqa
Note the strong correlation between $\dglb$ and $\dgrb$ ($0.9$) even
though $\xib$ and $\zetab$ were virtually uncorrelated ($-0.02$).  This
correlation stems from the fact that the error on $\zetab$ was so much
larger than that on $\xib$, as is evident from Fig.~\ref{fig3}.
Therefore, some care is necessary when using these limits.  For
instance, if we assume that $\dgrb = 0$, then the limits on $\dglb$
will be given by
\beq
\dglb = -0.0020 \pm 0.0011.
\eeq
and the central value of $\dglb$ changes sign!

It is clear from this analysis that $g^{b}_{R}$ is one of the least well
known of the precision electroweak observables.  Given that the
nominal standard model expectation is $\grb\sim 0.08$, the fractional
error on $\grb$ quoted in Eq.~(\ref{couplings}) amounts to roughly
15\%.  It would, of course, be of interest to find other measurements
that could be used to reduce this error.

\newpage

\section{Precision Electroweak Data and Supersymmetric $SU(5)$ Unification}

\begin{figure}[tp]
\centering
\unitlength=1cm

\begin{picture}(15,10)
\unitlength=1mm
\epsfbox[ 16 64 412 356]{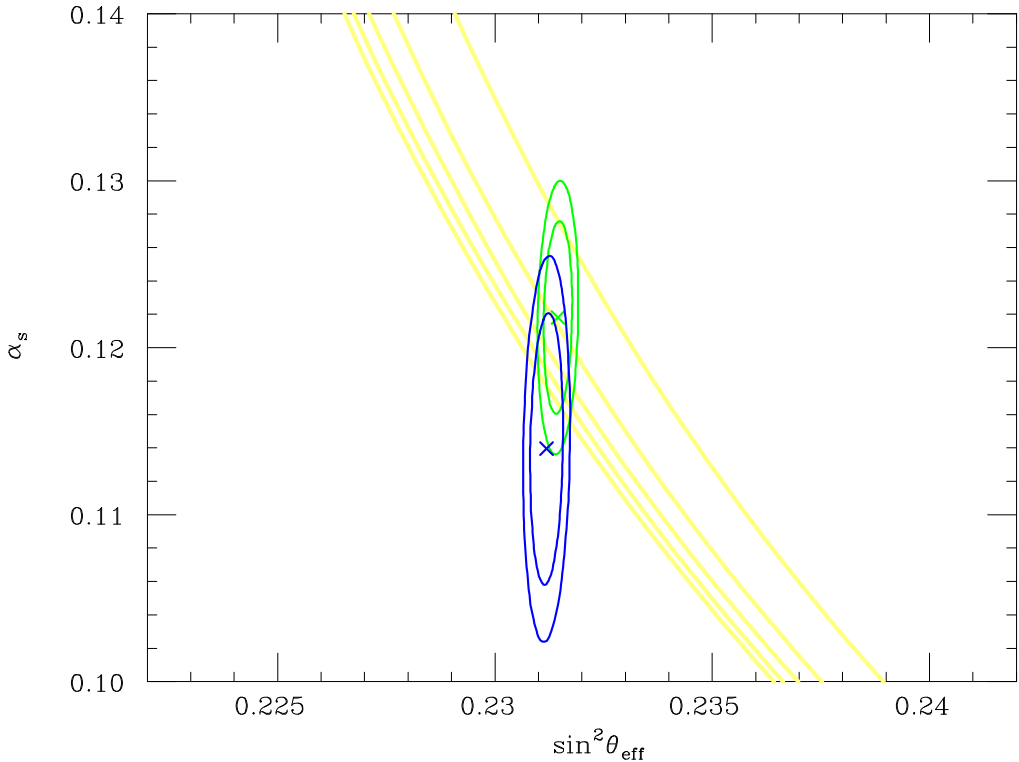}
\end{picture}

\caption[]{
$\sin^2\theta_\mathrm{eff}^\mathrm{lept}$ and $\alpha_s$.  Blue
contours: with non--zero $\xi_b$ and $\zeta_b$, Green contours: $\xi_b
= \zeta_b = 0$.  Yellow lines represent the 2--loop MSSM predictions
with all SUSY particle masses set to a common value
$M_\mathrm{SUSY}$.  The five lines correspond respectively to
$M_\mathrm{SUSY} = M_\mathrm{top}$, and 1, 2, 3, and 4~TeV from right
to left.  The thickness of the lines DOES NOT represent the
theoretical error which is much larger.  }
\label{fig4}
\end{figure}

We can also use the results of this analysis to assess the status of
supersymmetric grand unification \cite{DIMOPOULOS:81}.  From the
previous section, we can extract estimates for the values of $\seff$
and $\alpha_s(\mz)$:
\beqa
\seff         & = & 0.23118 \pm 0.00025  \cr
\alpha_s(\mz) & = & 0.114   \pm 0.005  
\eeqa
Note that allowing for non--zero $\xib$ and $\zetab$ has lowered the
preferred central values of both $\seff$ and $\alpha_s(\mz)$. It has
been noted \cite{GHILENCEA:98} that the prediction of $\alpha_s(\mz)$
from supersymmetric grand unification is somewhat high relative to
experiment.  If, as our analysis suggests, the value of $\seff$ is
somewhat smaller than previously thought, the GUT prediction of
$\alpha_s(\mz)$ will increase still further: A smaller value of
$\seff$ implies a smaller ratio of $g'/g$ at the $Z$ mass scale, and
this in turn increases the unification scale.  Since $\alpha_s(\mz)$
increases with the unification scale, a smaller value of $\seff$
implies a larger value of $\seff$.

In Fig.~\ref{fig4} we display the most naive prediction of
$\alpha_s(\mz)$ as a function of $\seff$ for a few values of the SUSY
mass scale, together with the 68 and 90 \% confidence level error
ellipses derived from our analysis.  We can see that the GUT
prediction of $\alpha_s(\mz)$ is slightly high relative to experiment
if $M_{\rm SUSY} \sim M_{\rm top}$ and the corrections from $\xib$ and
$\zetab$ are included.  We also display the result for
$\xib=\zetab=0$, for which the agreement is better. The GUT prediction
of $\alpha_s(\mz)$ can be lowered by threshold corrections to the
standard model couplings at the GUT scale. \cite{GHILENCEA:98}

More detailed analyses of the status of grand unification can be found
in the literature \cite{GHILENCEA:98}.  Our point here is simply to
note that the analysis of the previous section tends to shift
$\alpha_s(\mz)$  and $\seff$ away from the SUSY $SU(5)$ prediction, if
$\xib$ and $\zetab$ turn out to be non-zero.

\section{Conclusions}

We have reviewed the status of precision electroweak data using the
methods of Refs.~\cite{PESKIN:90,DPF:94} to parametrize potential
deviations from the standard model.  Agreement between the
standard model and experiment is quite good.  Indeed, all of the
parameters used in our analysis are found (for some choice of standard
model parameters) to be consistent with zero at the 90\% confidence
level, indicating good agreement between the minimal standard model
and experiment.

A few changes relative to previous analyses are apparent: First, the
apparent excess in $R_b$ reported in 1995 has decreased, and the
contraints on $U$ have improved substantially as knowledge of the $W$
mass has improved. The overall quality of the fit is improved if
either the Higgs mass is below our nominal value of 300~GeV, or if the
inverse fine structure constant $\alpha^{-1}(\mz)$ is somewhat larger
than our nominal value of 128.9.

\section*{Acknowledgements}

We would like to thank K.~S.~McFarland for providing us with
the latest NuTeV data.  This work was supported in part by the 
U.~S. Department of Energy, grant DOE-FG03-91ER40662, Task C.


\end{document}